\documentclass[12pt,a4]{article}
\usepackage{amsmath}
\usepackage{bm}
\usepackage{amsfonts}
\usepackage{amssymb}
\usepackage{graphics}
\usepackage[normal]{caption2}
\usepackage{subfigure}
\usepackage{rotating}
\usepackage{citesort}
\def\be{\begin{equation}}
\def\ee{\end{equation}}
\def\ba{\begin{array}}
\def\ea{\end{array}}
\def\bea{\begin{eqnarray}}
\def\eea{\end{eqnarray}}

\begin{document}
\baselineskip 20pt \setlength\tabcolsep{2.5mm}
\renewcommand\arraystretch{1.5}
\setlength{\abovecaptionskip}{0.1cm}
\setlength{\belowcaptionskip}{0.5cm}
\pagestyle{empty}
\newpage
\pagestyle{plain} \setcounter{page}{1} \setcounter{lofdepth}{2}
\begin{center} {\large\bf Isospin-dependent nucleon-nucleon cross section and symmetry energy: sensitivity towards collective transverse flow}\\
\vspace*{0.4cm}

{\bf Sakshi Gautam} and {\bf Rajeev K. Puri}\footnote{Email:~rkpuri@pu.ac.in}\\
{\it  Department of Physics, Panjab University, Chandigarh -160
014, India.\\}
\end{center}

\section*{Introduction}
The ultimate goal of studying isospin physics via heavy-ion
reactions with neutron-rich, stable or radioactive nuclei is to
explore the isospin dependence of in-medium nuclear effective
interactions and the equation of state of neutron-rich nuclear
matter, particular the isospin dependent term in the equation of
state. Because of its great importance to nuclear physics
community as well as to astrophysicists, significant progress has
been achieved by the establishment of existing and upcoming
radioactive ion beam facilities around the world \cite{rib}. The
collective transverse in-plane flow has been used extensively over
the past three decades to study the properties of hot and dense
nuclear matter, i.e. nuclear matter EOS and in-medium nucleon
nucleon (nn) cross section. The study of isospin effects helps us
to obtain the information about the isospin-dependent mean field.
The study of isospin effects in collective flow shows that the
isospin effects occur due to the competition between nn
collisions, symmetry energy, Coulomb potential \cite{li96,gaum}.
In the present work, we aim to see the relative contribution of
symmetry energy and isospin-dependent nn cross section towards the
collective transverse in-plane flow. The study is carried out
within the framework of isospin-dependent quantum molecular
dynamics (IQMD) model \cite{hart98}.

\section*{The model}
The IQMD model treats different charge states of nucleons, deltas,
and pions explicitly, as inherited from the
Vlasov-Uehling-Uhlenbeck (VUU) model. The isospin degree of
freedom enters into the calculations via symmetry potential, cross
sections, and Coulomb interaction.
  The nucleons of the target and projectile interact by two- and three-body Skyrme forces, Yukawa potential and Coulomb
 interactions.
 A symmetry potential between protons and neutrons
 corresponding to the Bethe-Weizsacker mass formula has also been included. The hadrons propagate using Hamilton equations of motion:
\begin {eqnarray}
\frac{d\vec{{r_{i}}}}{dt} = \frac{d\langle H
\rangle}{d\vec{p_{i}}};& & \frac{d\vec{p_{i}}}{dt} = -
\frac{d\langle H \rangle}{d\vec{r_{i}}}
\end {eqnarray}
 with
\begin {eqnarray}
\langle H\rangle& =&\langle T\rangle+\langle V \rangle
\nonumber\\
& =& \sum_{i}\frac{p^{2}_{i}}{2m_{i}} + \sum_{i}\sum_{j>i}\int
f_{i}(\vec{r},\vec{p},t)
 \nonumber\\ & & V^{ij}(\vec{r}~',\vec{r})
f_{j}(\vec{r~}',\vec{p~}',t) d\vec{r}~ d\vec{r~}' d\vec{p}~
d\vec{p~}'
 \nonumber\\
\end {eqnarray}
 The baryon potential\emph{ V$^{ij}$}, in the above relation, reads as
 \begin {eqnarray}
  \nonumber V^{ij}(\vec{r}~'-\vec{r})& =&V^{ij}_{Sky} + V^{ij}_{Yuk} +
  V^{ij}_{Coul} + V^{ij}_{sym}
    \nonumber\\
   & =& [t_{1}\delta(\vec{r}~'-\vec{r})+t_{2}\delta(\vec{r}~'-\vec{r})\rho^{\gamma-1}
   \nonumber\\
   &  & (\frac{\vec{r}~'+\vec{r}}{2})]
   +t_{3}\frac{\exp(|(\vec{r}~'-\vec{r})|/\mu)}{(|(\vec{r}~'-\vec{r})|/\mu)}
   \nonumber\\
   &  & +
    \frac{Z_{i}Z_{j}e^{2}}{|(\vec{r}~'-\vec{r})|}
   \nonumber\\
   &  & +t_{4}\frac{1}{\varrho_{0}}T_{3i}T_{3j}\delta(\vec{r_{i}}~'-\vec{r_{j}}).
 \end {eqnarray}
Here \emph{Z$_{i}$} and \emph{Z$_{j}$} denote the charges of
\emph{ith} and \emph{jth} baryon, and \emph{T$_{3i}$} and
\emph{T$_{3j}$} are their respective \emph{T$_{3}$} components
(i.e., $1/2$ for protons and $-1/2$ for neutrons).

\section*{Results and discussion}

  We simulate the reactions of Ca+Ca and Xe+Xe series having N/Z =
1.0, 1.6 and 2.0. In particular, we simulate the reactions of
$^{40}$Ca+$^{40}$Ca, $^{52}$Ca+$^{52}$Ca, $^{60}$Ca+$^{60}$Ca and
$^{110}$Xe+$^{110}$Xe, $^{140}$Xe+$^{140}$Xe and
$^{162}$Xe+$^{162}$Xe at impact parameter of
b/b$_{\textrm{max}}$=0.2-0.4. The incident energy is taken to be
100 MeV/nucleon.
\begin{figure}[!t] \centering
 \vskip -0.5cm
\includegraphics[width=15cm,height=12cm]{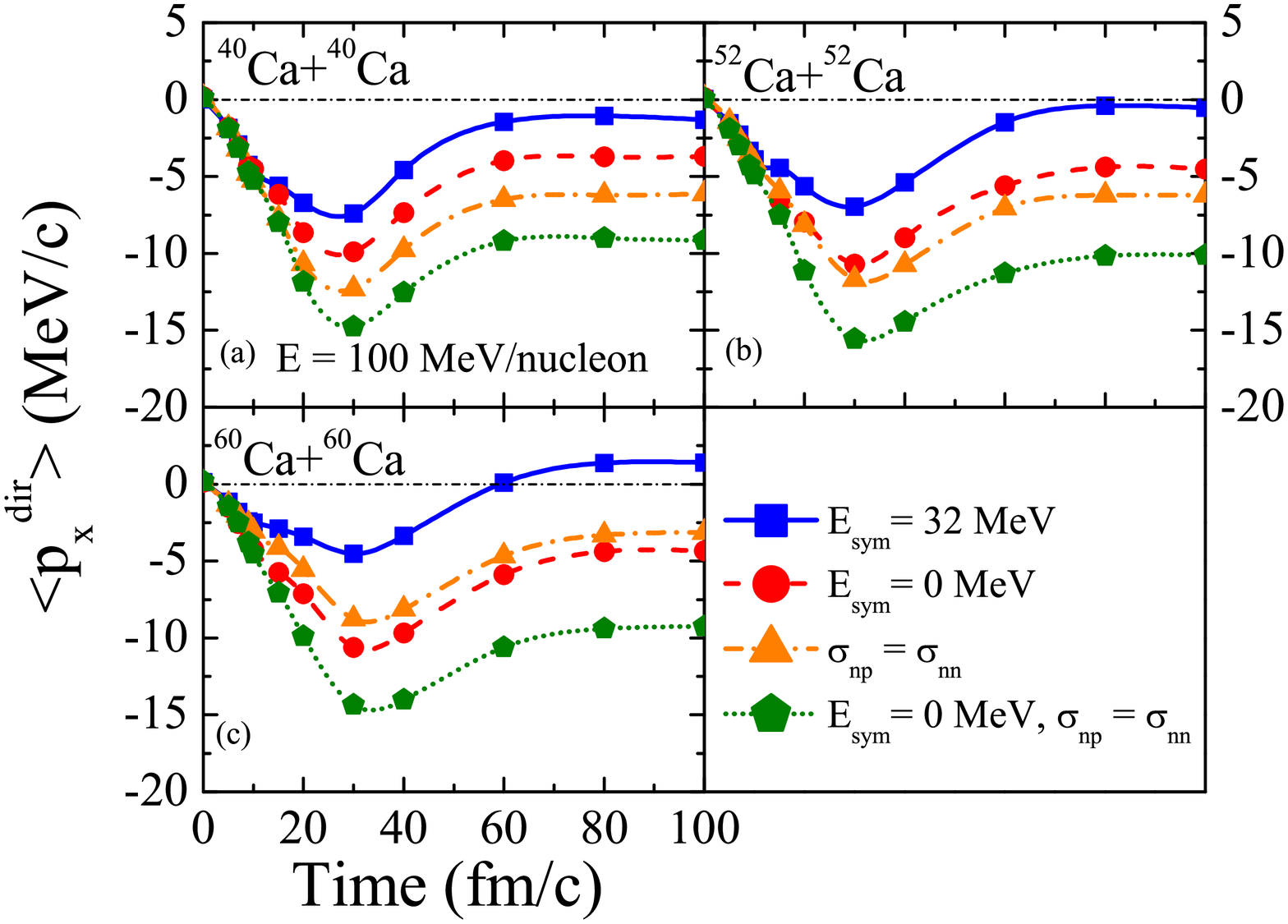}
 \vskip 0.5cm
\caption{ The time evolution of $<p_{x}^{\textrm{dir}}>$ for the
reactions of Ca+Ca having N/Z = 1.0, 1.6 and 2.0 at 100
MeV/nucleon. Lines are explained in the text.}\label{fig1}
\end{figure}

In fig. 1, we display the time evolution of
$<p_{x}^{\textrm{dir}}>$ for the reactions of $^{40}$Ca+$^{40}$Ca,
$^{52}$Ca+$^{52}$Ca and $^{60}$Ca+$^{60}$Ca. We see that at the
 start of the reaction, $<p_{x}^{\textrm{dir}}>$ (squares, solid lines) is
negative (due to the dominance of mean-field), reaches a minimum
and then increases and saturates at around 80 fm/c. The values of
$<p_{x}^{\textrm{dir}}>$ is maximum for higher N/Z reaction, i.e.,
$^{60}$Ca+$^{60}$Ca. Since we are having isotopes of Ca, so
Coulomb potential will be same for all the three N/Z reactions. So
the isospin effects in the collective flow will be due to the
interplay of symmetry energy and isospin-dependent nn cross
section. To see the effect of symmetry energy on the collective
transverse in-plane flow, we make the strength of symmetry energy
zero. The results are displayed by circles (dashed lines). We see
that when we make the strength of symmetry energy zero, the
collective transverse in-plane flow decreases in all the three
reactions. The decrease in flow is due to the fact that symmetry
energy is repulsive in nature and hence leads to positive in-plane
flow and so when we make it's strength zero, the flow decreases.
To see the effect of isospin dependence of nn cross section, we
make the cross section isospin independent, i.e.,
$\sigma_{np}=\sigma_{nn}$ and calculate the flow. The results are
displayed by triangles (dash-dotted lines). We find that the flow
decreases when we make the cross section isospin independent. This
is because in isospin dependent case, the neutron-proton cross
section is three times that of neutron-neutron or proton-proton
cross section. When we make the cross section isospin independent
the effective magnitude of nn cross section decreases which leads
to less transverse flow. Finally to see the combined effect of
symmetry energy and isospin dependence of cross section, we make
both the strength of symmetry energy zero and cross section to be
isospin independent, simultaneously. The results are displayed by
pentagons (short-dotted lines). We find that the maximum decrease
in flow is for $^{60}\textrm{Ca}$+$^{60}\textrm{Ca}$ which goes on
decreasing as we are moving to symmetric systems.

\section*{Acknowledgments}
 This work has been supported by a grant from Centre of Scientific
and Industrial Research (CSIR), Govt. of India.

\end{document}